\documentclass[useAMS,usenatbib]{mn2e}
\usepackage{epsfig}

\def\sval{Soviet Astron. Let}
\def\sva{Soviet Astr}
\def\aap{A\&A}
\def\mnras{MNRAS}
\def\apj{ApJ}

\def\cygx2{Cyg~X-2}
\def\ha{H$\alpha$}

\newcommand{\ie}{{\it i.e.,}}
\newcommand{\eg}{{\it e.g.,}}

\newcommand{\lsim}{\la}
\newcommand{\gsim}{\ga}

\title{Simultaneous optical and X-ray high speed photometry of Cyg~X-2}

\author[G. Dubus, B. Kern, A. A. Esin, R. E. Rutledge \& C. Martin]{G. Dubus$^{1}$\thanks{Present address: Laboratoire Leprince-Ringuet, CNRS/IN2P3, Ecole Polytechnique, F-91128, Palaiseau, France}, B. Kern$^{2}$, A. A. Esin$^{3}$, R. E. Rutledge$^{1}$ and C. Martin$^{2}$.\\
$^{1}$California Institute of Technology, MC 130-33, Pasadena, CA 91107\\
$^{2}$California Institute of Technology, MC 405-47, Pasadena, CA 91107\\
$^{3}$Harvey Mudd College, 301 E. 12th Street, Claremont, CA 91711}

\begin{document}
\date{Accepted for publication}
\maketitle

\begin{abstract}
  The X-ray emission from X-ray binaries may originate in flares 
  occurring when magnetic loops anchored in the disc
  reconnect. In analogy with our Sun, \ha\ emission should arise as
  the accelerated electrons thermalize in the optically emitting disc,
  perhaps leading to correlated variability between X-rays, \ha\ and
  the optical continuum. We present simultaneous X-ray and optical
  high speed photometry of the neutron star low-mass X-ray binary Cyg
  X-2 to search for such correlations. The highest time resolution
  achieved is 5~ms in white light and 100~ms with a 3~nm filter
  centred on \ha. We find power on timescales $\ga$ 100s (flickering)
  in optical with a total r.m.s. of a few \%, about an order of
  magnitude less than that seen in X-rays. We do not find
  significant correlations between the X-ray and optical fluxes on
  short timescales, hence cannot conclude whether magnetic flares
  contribute significantly to the optical emission.
\end{abstract}


\section{Introduction}

Observations of low mass X-ray binaries (LMXBs) show that strong
emission at X-ray energies $\lsim 100\,{\rm keV}$ is nearly ubiquitous
in these systems (e.g. \citealt{done}).  The standard optically thick,
geometrically thin accretion disc \citep{shakura} cannot explain X-ray
emission beyond a few keV.  Thus, the observed hard X-rays as
well as strong evidence for extended ``coronal'' emission in eclipsing
LMXBs (\citealt{vp2}) require more elaborate accretion flow models.
A popular scenario, inspired by the solar corona, suggests that the
cool Shakura \& Sunyaev disc is sandwiched between two layers of hot
plasma heated by reconnection of magnetic field loops anchored in the
disc \citep{galeev,haardt}. The presence of such a magnetically-heated
corona would not be surprising, considering that angular momentum in
discs is most likely transported outwards by turbulent magnetic
stresses (\citealt{balbus}).  In the context of this model the
observed X-ray spectra are interpreted as time-averaged emission from
multiple coronal flares, while the characteristic X-ray time
variability seen both in neutron star and black hole systems (\eg\
\citealt{vdk2}) is then related to a distribution of flare strengths
and durations.

In our Sun, bursts of hard X-ray emission from coronal flares are
accompanied by simultaneous \ha\ line emission when the beam of
electrons accelerated during reconnection is thermalised in the
photosphere (\eg\ \citealt{dulk}).  If the origin of X-ray emission is
similar to that in solar flares, correlations between \ha\ and X-ray
emission should occur on timescales ranging from the rise time of the
flares ($\sim 0.01$s) to their life time (either the dynamical time of
the accretion disc at the radius where a flare occurs, or the thermal
time needed to dissipate the energy of the wound-up magnetic field, up
to $\sim$ tens of seconds at the outer edge of the disc).  In the
optical spectra of LMXBs, \ha\ is the strongest line (EW $\sim$ 1--10
\AA, FWHM $\sim$ 1000--3000 km~s$^{-1}$) and its double-peaked
emission profile is an unambiguous signature of the accretion disc
\citep{vp2}.  The disc is optically thick, so the line has to be
produced in a thermally inverted layer, presumably located above the
photosphere.  However, if we pursue the analogy with the Sun, the
heating source of this disc chromosphere, be it coronal flares,
viscosity, reprocessing of X-rays or MHD waves, has yet to be
understood.

A significant fraction of the optical flux in persistent LMXBs comes
from the reprocessing in the disc of high energy photons produced in
the inner parts of the accretion disk, or in the NS boundary layer
(\citealt{vp2, obrien2}). This effect is most clearly demonstrated by
the optical echoes of NS X-ray bursts observed in some systems. A few
second delay between these optical echos and the X-ray bursts is
interpreted as the light travel time to the reprocessing site
\citep[\eg\ ][]{obrien2}. This delayed heating of the disk by
non-local radiation can in principle be distinguished from
quasi-instantaneous local heating by particles accelerated in flares
using high time resolution simultaneous X-ray and optical
observations.

While the X-ray timing properties of LMXBs are well established,
little is known about short-timescale non-orbital variations in the
optical.  On timescales $\gsim 1$~min flickering in white light
(characterised by erratic flux changes by factors $\lsim 2$) is a
common occurrence in both persistent and transient LMXBs.  The NS
LMXB Sco~X-1 shows intermittent correlated X-ray/optical flares
related to non-local reprocessing (\citealt{ilo80,petro}). LMC~X-2
also displays similar behaviour \citep{mcgowan}. Recent monitoring of
low-luminosity black hole LMXBs has revealed intriguing optical flares
on timescales of tens of seconds \citep{hynes2}.  On the other hand,
on timescales $\lsim 1$~min, the variability of the optical continuum
or \ha\ emission in LMXBs and its correlation with X-rays are still
largely unexplored; such studies are hampered by the faintness of
most systems, even when in outburst.  Only two LMXBs have well
established X-ray/optical continuum correlated behaviour on very short
timescales (down to a few ms): GX 339-44 (\citealt{motch}) and XTE
J1118+480 (\citealt{1118,hynes}). In both the variable optical flux
has been interpreted as synchrotron emission from energetic flares
(\citealt{fabian,dimatteo}).

\begin{table*}
\centering
\begin{minipage}{130mm}
\caption{Log of Cyg X-2 Optical Observations.}
\begin{tabular}{@{}lcrrrcrcrr@{}}
\hline
Date & Band & $\Delta$t & start & stop & Aper & S/N & samples & \# frames & samples \\
2001 & & ms & \multicolumn{2}{c}{hrs} & pixels &  &  per frame &  & kept \\
\hline
Aug, 2 & wl &   5 & 4.107  & 4.943  & 7$\times$7 & 17.2 & 47 & 1200 & 56237\\
Aug, 2 & \ha & 100 & 5.842  & 8.173  & 3$\times$3 & 5.3 & 48 & 1100 & 52573\\
Aug, 3 & \ha & 100 & 29.490 & 30.325 & 3$\times$3 & 4.3 & 48  & 400 & 19164\\
Aug, 3 & wl &  10 & 30.970 & 34.801 & 7$\times$7 & 10.5 & 47 & 4400 & 113572\\
Aug, 4 & wl &   5 & 53.211 & 54.193 & 7$\times$7 & 16.5 & 46 & 1200 & 50484\\
Aug, 4 & \ha & 100 & 54.759 & 58.980 & 3$\times$3 & 4.6 & 48 & 2000 & 63041\\
\hline
\end{tabular}

\medskip
\ha\ refers to a 3~nm filter centred on 656.2~nm, {\it wl} refers to
observations taken in white light; $\Delta t$ is the time resolution
of the data; start and stop times are in hours from MJD 52123.0 UT ;
Aper. is the size of the (square) aperture used for photometry in
pixels; S/N is the observed signal-to-noise of each $\Delta t$
exposure; samples per frame is the number of exposures on each CCD
frame; last column is the number of flux measurements kept after bad
data points were removed (see \S2.2).
\end{minipage}
\end{table*}

\begin{figure*}
\centerline{\epsfig{file=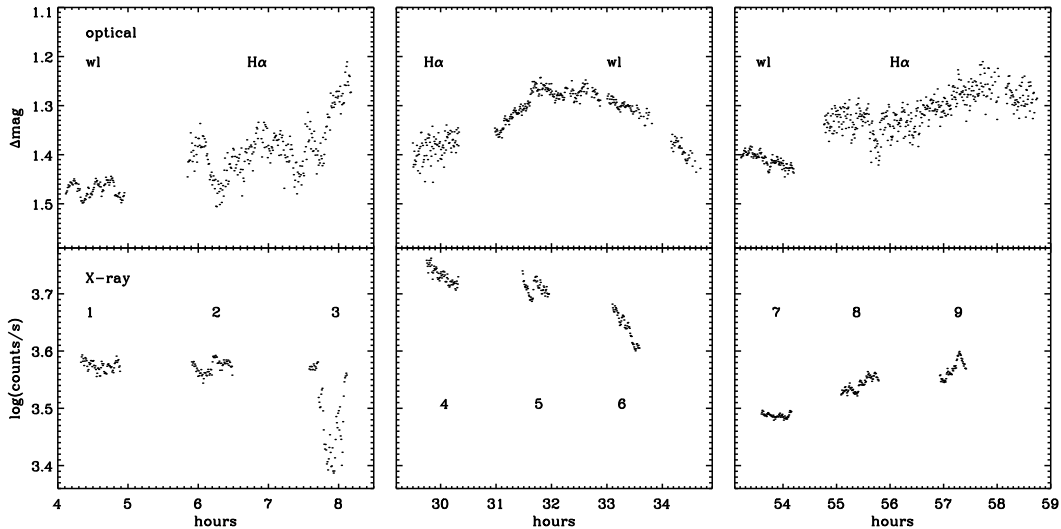}}
\caption{Simultaneous optical (top, Palomar 200-inch) and X-ray
(bottom, {\em RXTE}) lightcurves of \cygx2. The time is in hours from
MJD 52123. Each point is a 32 s (elapsed time) average. The type of
photometry (white light or \ha\ filter) is indicated above each
optical lightcurve. $\Delta$mag is the differential magnitude between
\cygx2 and the comparison star. The X-ray count rate was derived from
PCA channels 0--23 (2.0--9.8~keV). \cygx2 was on the normal branch
during the first two X-ray visits, on the flaring branch on the third
and on the horizontal branch in the subsequent observations. Typical
error bars for the 32 s photometry bins are 0.002-0.003 mag in white
light, 0.014-0.018 mag in \ha\ and 1.2-1.5$\cdot 10^{-3}$
log(counts/s) in X-rays.}
\end{figure*}

In this paper we report on simultaneous X-ray and \ha/white light high
time resolution observations of \cygx2, an optically bright ($V\sim$
14.8) persistent LMXB with a neutron star primary (\eg\
\citealt{ok}). \cygx2 has rapid (hour--days) X-ray spectral variations
closely related to the timing properties: the slope/amplitude of the
power law and the frequency of quasi-periodic oscillations (QPOs,
$\nu\sim$1-20~Hz) seen in the power spectrum vary with the X-ray
colours (\eg\ \citealt{wijnands}).  The optical flux varies by $\lsim
1$~mag on timescale of hours to days, and flickering is reported in
the brightest state.  An ellipsoidal modulation on the orbital period
of 9.8 days is observed with a $V$ band amplitude of 0.1~mag. The
donor star contributes about 50--70\% of the total optical flux. \ha\
is known to vary on timescales of hours to days (see \S4), but no
systematic study of the \ha\ line variability has ever been
undertaken. Here, we describe our search for correlated rapid
variations between the X-ray and optical fluxes of \cygx2.

\section{Observations and data reduction}

We studied the variability of the optical continuum, \ha\ line
emission and X-ray flux of \cygx2 at frequencies $\lsim 100\,{\rm Hz}$
using simultaneous data obtained on 2--4 August 2001 (UT) at the Hale
200-inch at Palomar and with the {\em Rossi X-ray Timing Explorer}
({\em RXTE}).

\subsection{X-ray data}

Cyg X-2 was observed with {\em RXTE/PCA} (\citealt{pca}) on three
consecutive nights (2--4 Aug 2001), for three consecutive orbits on
each night. Data were analysed using FTOOLS v5.0.  For all nine
separate observations, we used data types collected by the Experiment
Data System (EDS) in three modes ({\tt SB\_125us\_0\_13\_1s,
SB\_125us\_14\_17\_1s, SB\_125us\_18\_23\_1s}), which collected counts
of 2.0-5.7 keV (centroid energies), 6.1-7.4 keV, and 7.8-9.8 keV.  We
made use of event mode data ({\tt E\_125\_64M\_24\_1s}) in the 9.5-20
keV energy range.  The time-tags associated with the counts were
corrected to the solar-system barycenter in the DE405 time system
using {\tt axbary}.

We extracted a spectrum from the Standard 2 EDS mode data from the
first night's observation.  Examination of this spectrum showed that
the countrate in excess of the background by a factor of 2-3 orders of
magnitude up to 10 keV, and by a factor of 2-10 in the 10-20 keV
energy range.  We neglected the background for the spectral analyses
herein. The 2.0-9.8~keV lightcurves (rebinned on 32 s from the original
resolution of 1/256s) for each of the 9 visits are shown in the bottom
panel of Fig.~1. The dip in countrate during the third visit is
typical of the flaring branch (see \S3.1).

\subsection{Optical data}
High-speed optical photometry of \cygx2\ was obtained using a CCD in a
multiple frame-transfer observing mode.  The detector used is a SITe
SI-502A thinned, AR-coated, back-illuminated 512$\times$512-pixel CCD.
The F/9 Cassegrain focus of the Palomar 200-inch telescope illuminated
a 200$\times$4 arcsec slit, which was re-imaged onto the uppermost
rows of the CCD.  The image was demagnified to a plate scale of 0.4
arcsec/pixel, yielding a slit size of 500$\times$10 pixels.  The
un-illuminated portion of the CCD was used as storage, allowing 50
images to be accumulated before reading out the entire CCD.  The
transfer time, the time required to shift an image from the
illuminated region of the CCD into storage, is short, so the shutter
remained open during the accumulation of all 50 images, and was only
closed during readout.

The timing of the images in a single frame was controlled by a GPS
receiver and time-code generator.  The GPS timing allowed control of
the start time (and duration) of each image to an accuracy of 1
$\mu$s.  The transfer time is 420 $\mu$s (42 $\mu$s per row shift),
which occurs immediately following the start of a given image.  The
effective exposure time is the same for every image on the CCD except
for the first and last few images.  The first image or images have an
effective exposure time that is reduced by the time it takes the
shutter to open, which is approximately 13 ms.  The last image has an
effective exposure time that is increased by the time it takes the
shutter to close, although the readout takes place during this time,
resulting in some smearing.  For this reason, the first image or
images and the last image are discarded in the photometric analysis.

CCD pixels can be digitized in 22.4 $\mu$s, or discarded in 2.4
$\mu$s.  To minimize the time spent reading the CCD, two windows are
defined, which mark ranges of columns in which the pixels are
digitized, with pixels in all other columns discarded.  Each window
was 40 pixels wide, which allowed the entire CCD to be read out in 1.5
s.  Overhead in the computer operating system added up to one second
to the total time required for each frame.  The exposure times used in
these observations were 5 ms, 10 ms, and 100 ms per image, giving duty
cycles of approximately 10\%, 20\%, and 65\%, respectively.

The long slit, with two windows defined, ensured that each image
included a bright comparison star.  The comparison star used is the
photometric standard 35 of \citealt{hh97} ($V=13.498\pm 0.001$,
$B-V=1.194\pm 0.004$) situated $\sim$ 1\arcmin\ NNW of \cygx2.  This
star has been extensively monitored, and has shown no intrinsic
variability down to 0.001 mag. Roughly half of the data was taken with
a 3~nm FWHM filter centred on \ha\ and the remainder in white light
(see Tab.~1).  The seeing over three nights ranged from 0.8 to 1.2
arcsec FWHM in \ha, and from 1.2 to 1.6 arcsec FWHM in white light.

Lightcurves were constructed for each observation sequence using
square photometric apertures ranging in size from 1$\times$1 to
11$\times$11 pixels (0.4$\times$0.4 to 4.4$\times$4.4 arcsec). While
the use of square photometric apertures, rather than circular
apertures, is not conventional, there is little difference for
apertures composed of a small number of pixels.  For the larger
apertures, there was no measurable difference in S/N when circular
apertures were used, so for consistency, all photometric apertures
used were square.

For each image the aperture is placed at the centroid of \cygx2, known
to a fraction of a pixel. The comparison star aperture is placed at
the (fixed) mean distance from \cygx2\ to compensate for aperture
losses. We checked that the PSFs of the two stars were similar and not
affected by instrumental effects. Because every image includes both
\cygx2\ and the nearby comparison star, observed simultaneously under
identical observing conditions, differential photometry allows the
removal of temporal photometric throughput variations. We do not
correct for extinction hence the broad band white light data can be
subject to an airmass-correlated variation due to the colour difference
between \cygx2\ and the comparison star.  However, since all our
observations were done at airmass $<$1.2, with two-thirds of the data
taken at airmass $<$ 1.1, we feel this will have a negligible
influence on the short timescale variability study.

We ignored images for which the centroiding algorithm had failed, a
cosmic ray had been tagged inside the aperture, or (more commonly)
those images in which the photometric aperture extended outside the
unvignetted slit edges.  The background in each window of each image
was calculated as an average over the window, for pixels more than 10
pixels (4 arcsec) from the window center. We then computed the
signal-to-noise ratio of the flux from \cygx2 as a function of
aperture.  The maximum S/N was obtained for an aperture of 7$\times$7
pixels in white light and 3$\times$3 pixels with the \ha\ filter.  The
comparison star data was extracted using the same aperture and the
flux ratio computed to correct for lost flux and atmospheric
variations.  The final lightcurve is therefore the differential
magnitude between \cygx2 and the comparison star. The S/N, chosen
aperture, total number of measurements and number of measurements kept
in each observation sequence can be found in Tab.~1.

The lightcurves, rebinned with 32 s resolution, are shown in Fig.~1.
Note that 32 s is an elapsed time and that the actual exposure time and
S/N in each bin varies with the observational setup.  The 1$\sigma$
error for each bin is obtained from the standard deviation of the
averaged data points, hence implicitly assuming that the measurements
are independent. Typical error values are given in column 4 of Tab.~2.

\section{Analysis and Results}

\subsection{X-ray data}
To characterise the overall behaviour of \cygx2 in X-rays, a
colour-intensity diagram was plotted using 32 s flux averages
(Fig.~2).  This diagram suggests that \cygx2 was on the normal branch
during visits 1 and 2 (Aug~2), on the flaring branch during visit 3
(Aug~2) and on the horizontal branch during the visits of Aug~3 and 4.
Power density spectra (PDS, see \citealt{vdk}) were then constructed
for each visit using all the detected photons.  The high frequency
(1~Hz--128~Hz) part of the PDS is the average PDS of 1~s long segments
with a binsize of 1/256~s.  The low frequency part ($\sim
10^{-4}$--0.5~Hz) is computed from a 1~s rebinned lightcurve.  The PDS
are normalised to units of r.m.s.$^2$ Hz$^{-1}$, \ie\ fractional
variance per unit frequency, and the expected Poisson measurement
noise bias was subtracted.  No corrections were made for background
and PCA dead time, which are negligible.

The resulting PDS confirm that \cygx2 moved through the different
branches of the Z track during the observations.  On Aug~2, the
absence of detectable broad band variability at frequencies above 1~Hz
except for a weak QPO at 6$\pm$1~Hz during the second visit
(1.3$\pm0.2$\% r.m.s.) is typical of the normal/flaring branch.  The
Aug~3--4 PDS are typical of the horizontal branch with strong broad
band variability peaking around 10~Hz and two QPOs varying between
$\sim$24--34~Hz (3.1$\pm$0.2\% to 4.4$\pm$0.2\% r.m.s.) and
$\sim$50--63~Hz (2.2$\pm$0.3\% to 2.3$\pm$0.3\% r.m.s.).  In addition,
low frequency variability is present in all the observations at
frequencies below 0.1~Hz.  These results are in line with previous,
more extensive X-ray timing studies of this source (e.g.
\citealt{wijnands,kuulkers,w2001}).  We show in Fig.~3 some of the
logarithmically rebinned X-ray PDS obtained on different nights.

\begin{figure}
\centerline{\epsfig{file=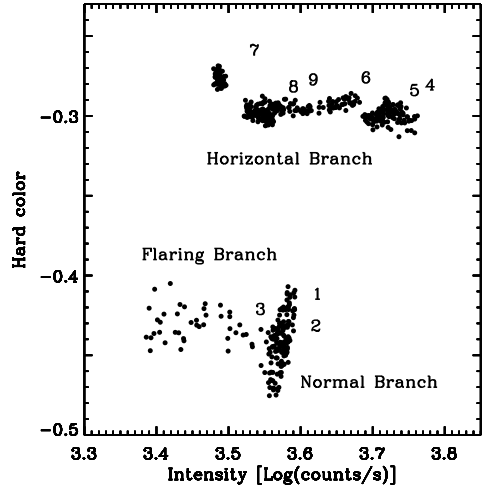}}
\caption{Hardness-intensity diagram of \cygx2 on Aug. 2, 3 and 4. The
hard colour (y-axis) is the log of the count rate ratio between
5.7-9.8 and 2.0-5.7~keV (PCA channels 14-23 and 0-13). The intensity
(x-axis) is from 2.0 to 9.8~keV. Each point is a 32~s average. The
numbers indicate the position of the points from each X-ray visit (as
labelled in Fig.~1).}
\end{figure}

\subsection{Optical data}

The 32 s binned optical lightcurves (Fig.~1) show variations above the
measurement noise on timescales of minutes to hours.  Quantitatively,
the binned white light (resp. \ha) lightcurves have standard
deviations of up to 0.04 mag (resp. 0.06) compared to the measurement
noise of $\sim$ 0.003 mag (resp. 0.017; see \S2.2).  The values for
each observation are in Tab.~2.  Plotting the optical fluxes against
the simultaneous X-ray fluxes, we find no obvious correlations between
the two; but we note that the average white light flux was highest (by
about 0.15 mag, second night), when the X-ray count rate was also
highest (close to the vertex of the horizontal branch).

PDS were constructed for the optical data using a slight modification
to the standard technique used for X-ray timing.  For the optical
data, we found the white noise level empirically instead of using
Poisson statistics.  The data were (fast) Fourier transformed and
normalised by multiplying the amplitudes by 2/Var($f$), where Var($f$)
is total variance of the data $f$.  This ensures the powers follow a
$\chi^2_2$ distribution when the lightcurve is dominated by
(measurement) white noise (\ie\ each point is independent and randomly
taken from the initial flux distribution).  The optical power spectra
typically showed power at low frequencies and white noise at high
frequencies.  In these cases we divided by the average power at the
frequencies where measurement noise dominates to recover the
$\chi^2_2$ statistics.  The Fourier amplitudes give the fractional
variance per frequency bin and, after subtraction of the constant
white noise component, we plot the power $\times$ frequency in
(dimensionless) units of r.m.s.$^2$ to highlight the timescales with
significant variability and ease comparison with the X-ray PDS
(Fig.~3).

The low duty cycle of the data acquisition scheme introduces large
amounts of spurious power in a Fourier transform covering several
frames.  To circumvent this problem, we separated the calculation of
the power spectrum into two parts so that the transformed lightcurves
had duty cycles close to 100\%.  At high frequencies, we computed a
power spectrum of the images in each frame separately, then averaged
the power spectra of all the frames.  Since there are between 46 to 48
images (photometry points) in each frame with a time resolution of
5~ms (white light) to 100~ms (\ha), the resulting frequency range is
about 5--100~Hz (white light) or 0.2--4.8~Hz (\ha), depending upon the
observational setup (see Tab.~2).  At lower frequencies we took the
average flux of each frame (the sum of all images in a frame) and
computed the power spectrum of the resulting lightcurve.  The
resulting frequency range is about $10^{-3}$--0.05~Hz (see Tab.~2).
The combined power spectrum is free of power leakage at the expense of
a loss in frequency resolution and coverage.  At high frequencies, the
averaging of the individual spectra diminish the measured power from a
highly coherent signal, because the contributions from each frame are
not summed in-phase.

Significant power is present in all of the datasets at very low
frequencies (below $3\times 10^{-3}$~Hz in white light and
$10^{-3}$~Hz in \ha) except during the short August 3 \ha\
observation. The total r.m.s. of the flickering is estimated by
integrating over the frequencies where power is detected above the
3$\sigma$ single trial level.  We obtain values of 1--3\% r.m.s. (see
Tab.~2) which are consistent with the standard deviation of the 32 s
binned lightcurve.  The optical variability is about an order of
magnitude smaller than that seen in the X-ray lightcurves on the same
frequency range.

At higher frequencies, statistically significant power is detected in
the white light observation of Aug.~3, when \cygx2 was on the
horizontal branch with a high X-ray flux.  The total r.m.s. is about
1\% (0.02--0.07~Hz). Unfortunately, this is the only detection of
optical variability at high frequency in our limited
dataset. Independent confirmation would be desirable. The very low
frequency power in both optical and X-rays is reasonably approximated
by $P_\nu \sim \nu^{-2}$ (\ie\ $\nu P_\nu \sim \nu^{-1}$ in Fig.~3).
We obtained an upper limit to the flickering power in the cases where
none was detected by calculating the total r.m.s. of power with this
particular shape that would have yielded a 3$\sigma$ detection
(typically a few \%, see Tab.~2).  We also computed 3$\sigma$ upper
limits to the r.m.s. of a sinusoidal signal.  Both of these upper
limits take into account the number of trials in the frequency range.
All of the results may be found in Tab.~2.

\begin{table*}
\centering
\begin{minipage}{140mm}
\caption{Optical Variability of \cygx2}
\begin{tabular}{@{}lclcllclrrclrr@{}}
\hline
Date & Band & $\Delta$t & & err$^\star$ & $\sigma^\star$ & & $\nu_{\rm min}$-$\nu_{\rm max}$ & pulse & total & & $\nu_{\rm min}$-$\nu_{\rm max}$ & pulse & total \\
 & & ms & & \multicolumn{2}{c}{10$^{-3}$ mag} & & \multicolumn{1}{c}{mHz} & \multicolumn{2}{c}{\% r.m.s.} & & \multicolumn{1}{c}{Hz} & \multicolumn{2}{c}{\% r.m.s.}\\
\hline
Aug, 2 & \ha & 100 & & 14 & 57 & & 0.12-66 & $<$0.4 & 3.4 & & 0.21-5 & $<$1.2 & $<$2.5 \\
Aug, 3 & \ha & 100 & & 18 & 24 & & 0.30-66 & $<$0.7 & $<$1.6 & & 0.21-5 & $<$1.9 & $<$3.8 \\
Aug, 4 & \ha & 100 & & 16 & 37 & & 0.07-66 & $<$0.5 & 2.1 & & 0.21-5 & $<$1.4 & $<$2.8 \\
\hline
Aug, 2 & wl &  5 & & 2.4 & 14 & & 0.33-200& $<$0.2 & 1.0 & & 4.35-100 & $<$0.3 & $<$0.6 \\
Aug, 3 & wl & 10 & & 2.0 & 41 & & 0.07-160& $<$0.2 & 2.2 & & 2.12-50 & $<$0.3 & 1.1 \\
Aug, 4 & wl &  5 & & 2.6 & 14 & & 0.28-160& $<$0.2 & 0.8 & & 4.35-100 & $<$0.4 & $<$0.7 \\
\hline
\end{tabular}

\medskip
{\em wl} is white light; err$^\star$ is the estimated error of a 32 s
photometry bin (in mmag) and $\sigma^\star$ is the standard deviation
of the 32 s binned lightcurve (in mmag; see Fig.~1); the optical
variability is studied in two frequency ranges (see \S3.2 for
details): $\nu_{\rm min}$-$\nu_{\rm max}$ give the corresponding
frequency ranges; pulse gives the r.m.s. upper limit for a sine
signal; total is the broad band r.m.s. variability or upper limit
(assuming $F_\nu \propto \nu^{-2}$ coloured noise).
\end{minipage}
\end{table*}

\begin{figure*}
\centerline{\epsfig{file=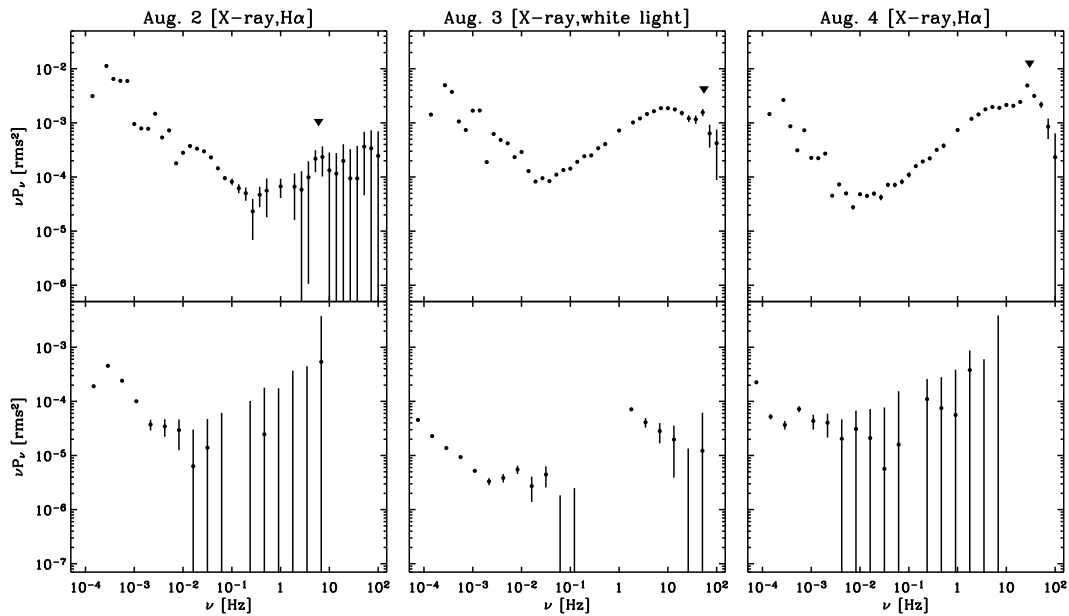}}
\caption{Temporal variability power spectra for the X-ray (top) and
optical (bottom) data on different nights. White noise has been
removed, the power normalised to units of r.m.s.$^2$~Hz$^{-1}$ and
multiplied by frequency. The frequency range that contributes most to
variability and its r.m.s. are thus easily identified. We also plot
3$\sigma$ (single trial) error bars to help assess the statistical
significance of the various features. The bottom of the first panel
shows the power spectrum of the long \ha\ dataset taken on August 2;
top shows the X-ray power spectrum from the simultaneous RXTE data
(two visits, see Fig. ~1). Second (resp. third) panel is for the long
white light (resp. \ha) dataset taken on August 3 (resp. 4). The power
spectra have been logarithmically rebinned. Downward arrows in the
X-ray panels highlight the location of the QPOs discussed in \S3.1.}
\end{figure*}

\subsection{Cross-correlation}

We cross-correlated each of the nine simultaneous X-ray and optical
datasets (Tab.~1).  The X-ray data, which have the highest time
resolution, were corrected to the local Palomar time and rebinned in
an identical way to the optical data set, yielding two continuous
lightcurves with the same resolution ($\Delta t=$ 5, 10 or 100ms).
The cross-correlation function at a lag $k\Delta t$ is defined as
(e.g. \citealt{krolik}):
\begin{equation}
C(k\Delta t)=\frac{1}{M}\sum\limits_{j-i=k} \frac{(o_i-\bar{o})(x_j-\bar{x})}{\sigma_{\rm X}\sigma_{\rm opt}},
\end{equation}
where $o$ is the optical lightcurve, $x$ the X-ray, the bar symbolises
the average of the dataset and $\sigma$ are the standard deviations.
$C(k\Delta t)$ is an average over the $M$ pairs of points which are
$k\Delta t$ apart.  The variance of the $M$ averaged discrete
correlations gives an estimate of the error on $C$.  The above is
efficiently computed in Fourier space for continuous lightcurves with
the same $\Delta t$:
\begin{equation}
C(k\Delta t)=\frac{1}{\sigma_{\rm X}\sigma_{\rm opt}} \left\{ \frac{{\rm F}^{-1} ({\rm F}_o {\rm F}_x^\star)} {{\rm F}^{-1}({\rm F}_{w_o} {\rm F}_{w_x}^\star)}\right\}_k,
\end{equation}
where $\rm F$ is the (fast) Fourier transform, $\star$ is the
conjugate and the $w$ are window functions (a timeseries with value 1
when data are taken, 0 elsewhere).

There were no obvious trends on timescales $\gsim$10~min where the
datasets are not long enough to clearly establish correlations.  For
the cross-correlations below 10~min, we subtracted piecewise linear
fits in 600~s increments to the datasets to smooth out any underlying
long timescale variation.  The X-ray and optical data did not show any
significant correlations down to 5~ms (100~ms). Cross-correlating with
different energy selected X-ray lightcurves (2.0--5.7 keV, 5.7--9.8
keV, 9.8 keV and above) gave identical results.

\section{Discussion}
Our dataset is consistent with previous studies reporting variations
in the optical continuum of \cygx2 of $\sim$0.5~mag in days,
$\sim$0.1~mag in hours and $\sim$0.05~mag in minutes
(\citealt{kristian,kru74,kilyachkov,beskin}). Orbital ellipsoidal
modulation aside, there are few well-established features in the
optical variability of \cygx2. There is a trend for the continuum to
become bluer in $U$-$B$ as \cygx2 becomes brighter ($B$-$V$ staying $\sim$
constant; \citealt{lyutyi,basko}). Varying fluxes and FWHM on timescales
$\gsim$ 1 hr were also noticed early on in H$\alpha$, H$\beta$, He{\sc
  ii} $\lambda$4686 and the Bowen blend (\citealt{johnson,cowley}).

The evolution from the horizontal to the flaring branches in Cyg X-2
like sources has been proposed to reflect an increasing mass accretion
rate in the disc (see discussion in \citealt{homan}). This could be
due to fluctuations in the mass transfer rate from the donor star or
to fluctuations in the angular momentum transport processes,
presumably on the disc viscous timescale. Interestingly, \ha\ is
sometimes in absorption (\citealt{chuvaev}): in dwarf novae, the
Balmer lines evolve from emission to absorption during the rise to
outburst when the mass accretion rate increases
(e.g. \citealt{szkody}). However, mass accretion rate changes are
hard to reconcile with rapid (hours) variations changes unless they
originate in the inner disc where the timescales are short.

Variations in the inner region can affect the optical output of
the disc via non-local irradiation. The observed correlation between
high ionisation line fluxes (He{\sc ii}, C{\sc iii}/N{\sc iii}) and
the optical continuum of \cygx2 does suggests that reprocessing plays
a major role \citep{ilo79,vp,obrien}. Some of the long timescale X-ray
fluctuations of \cygx2 have been argued to be due to varying
obscuration, perhaps by an inner disc warp
(\citealt{kvdk,kvdkv,vrtilek,vrtilek2}). Changes in the albedo or
geometry of reprocessing would lead to complex correlations between
X-rays and optical \citep{esin}. O'Brien (2001) found that the optical
and X-ray fluxes were correlated on the horizontal branch, but
anticorrelated on long timescales in the normal and flaring branches.
Curiously, on shorter timescales ($\sim$1~min), the X-ray and optical
on the flaring branch were correlated. O'Brien proposed that
several components with different intrinsic timescales contribute to
the optical emission. Our observation of an order of magnitude lower
variability in optical compared to X-rays on long timescales (\S3.2)
is consistent with reprocessing of a fraction of the X-rays, but we
did not find significant correlations to support this.

As argued in \S1, variability and X-ray/optical correlations on
short timescales could provide constraints on emission sites in
\cygx2. Previous studies found no variability in the optical continuum
below $\sim 1$~min (\citealt{che78,aur78,ilo78}). With a factor 2--5
better sensitivity and a longer dataset, we find some evidence for
broad band variability {at frequencies of few Hz in the continuum
when \cygx2 was optically bright and on the horizontal branch; but we
find no correlations with the X-ray noise component that peaks at
$\sim 10$~Hz in this state. For rapid line variability,
\citet{canizares} give an upper limit of 8\% to the pulsed fraction of
the He{\sc ii} flux between 0.1 and 125~Hz.  We do not find
variability in the \ha\ line on timescales shorter than 1~min.

In contrast with \cygx2, the other optically bright Z source, Sco X-1,
shows correlated X-ray and optical flares on a 1--60~s timescale when it
is on the flaring branch (\citealt{ilo80,petro}). The optical flares
are 20 s filtered versions of the X-rays, probably as a result of
non-local reprocessing. Similar behaviour is not seen in \cygx2,
although the X-ray flux and variability are comparable. A different
irradiation geometry, as mentioned in the previous paragraph, might
explain their absence.

We did not find evidence for correlated X-ray -- \ha\ variations due
to solar-like flares. However, \ha\ emission could be powered by
flares that are too weak to be detected individually and occurring at
a fast enough rate that correlations are lost. For instance,
\citet{poutanen} discuss a model for the X-ray timing characteristics
of Cyg X-1 in which the rapid variability is due to flare avalanches
in the disc. Only rare large flares from the high luminosity end of
the distribution function can be distinguished in the X-ray flux of
Cyg X-1 \citep{gierlinski}.

Calculating the thin disc response to irradiation by high energy
electrons produced in flares, \citet{williams} find that an avalanche
model can explain the H$\beta$ emission and the optical continuum
flickering observed in cataclysmic variables (which have accretion
discs comparable to those in LMXBs). Only 4\% of the flare energy is
used to produce the variable $V$ band flux. A similar model might
apply to \cygx2 with, perhaps, more of the flare energy deposited in
the disc to take into account the higher optical/X-ray flux ratio.
The high number of active regions required} is likely to have a
significant impact on the X-ray spectral properties (\eg\
\citealt{dimatteo2}). X-ray emission by bremsstrahlung or Compton
upscattering is not modelled by \citet{williams}. If flares give
rise to the broad band X-ray noise (with r.m.s. levels of a few \%)
then our upper limit of $\sim 3$\% r.m.s is too high to detect their
\ha\ signature.

\section{CONCLUSION}
We obtained \ha\ or white light lightcurves of \cygx2 with high
time resolution simultaneously with X-ray data. Two (mutually non
exclusive) mechanisms may link the two wave bands: non-local
reprocessing of X-ray photons emitted close to the compact object or
local reprocessing of particles accelerated in magnetic flares. We
find broad band variability on timescales longer than a minute and
some evidence on sub-second timescales in one white light
observation. No correlations are found. Our observations cannot be
used to distinguish between non-local or local reprocessing. The lack
of correlations could be explained either by unfavourable irradiation
circumstances (non-local case) or because large, identifiable magnetic
flares are scarce (local case).

The prospects offered by rapid optical photometry have proven
difficult to fulfill. Part of the difficulty lays in assembling a
large, coherent, simultaneous X-ray and optical dataset with secure
calibrations from which trends can emerge (\ie\ the strategy behind
the success of X-ray timing studies). More problematic are the small
amplitude of the expected variability and limited number of bright
enough objects for which sensitive enough observations and comparisons
can be made. In principle, the present generation of 8--10 m class
telescopes could remedy that.

\section*{acknowledgments}
We thank Jean Swank for approving the observation for simultaneous
observing with Palomar 5m. This work was supported by a Chandra
Postdoctoral Fellowship grant \#PF8-10002 awarded by the Chandra X-Ray
Center, which is operated by the SAO for NASA under contract
NAS8-39073.

\end{document}